%% file: hyades.tex
\documentclass{aa}
\usepackage{psfig}

\begin{document}

\newcommand{\al}  {\rm et al.}
\newcommand{\eg}    {\em e.g.}
\newcommand{\ie}    {\em i.e.}
\newcommand{\met}   {metallicity }
\newcommand{\kms}   {km.s$^{-1}$}
\newcommand\beq{\begin{equation}}
\newcommand\eeq{\end{equation}}
\newcommand\beqa{\begin{eqnarray}}
\newcommand\eeqa{\end{eqnarray}}
\newcommand{\aaa} [2]{A\&A {\bf #1}, #2}
\newcommand{\aas} [2]{A\&A Suppl. {\bf #1}, #2}
\newcommand{\aj}  [2]{AJ {\bf #1}, #2}
\newcommand{\apj} [2]{Ap. J. {\bf #1}, #2}
\newcommand{\apjl}[2]{Ap. J. Letter {\bf #1}, #2}
\newcommand{\apjs}[2]{Ap. J. Suppl. {\bf #1}, #2}
\newcommand{\araa}[2]{A\&AR {\bf #1}, #2}
\newcommand{\pasp}[2]{PASP {\bf #1}, #2}
\newcommand{\mnras}[2]{MNRAS {\bf #1}, #2}

\thesaurus{10.15.2;        
           08 (08.06.3;    
               08.02.3;    
               08.05.3;    
               08.08.1;    
               08.19.1)    
          }

\title{Consequences of Hipparcos parallaxes for stellar evolutionary models$^{\star}$.}  
\subtitle{Three Hyades binaries: V818 Tauri, 51 Tauri, and $\theta^2$ Tauri}

\author{Erwan Lastennet \inst{1,2}, David Valls--Gabaud \inst{2,3}, 
Thibault Lejeune \inst{4} and Edouard Oblak \inst{5}}

\institute{Astronomy Unit, Queen Mary and Westfield College,
           Mile End Road, London E1 4NS, UK
     \and  UMR CNRS 7550, Observatoire Astronomique,
           11, rue de l'Universit\'e, 67000 Strasbourg, France
     \and  Institute of Astronomy, Madingley Road,
           Cambridge CB3 0HA, UK
     \and  Astronomisches Institut der Universit\"at Basel, Venusstr. 7, 
           CH-4102 Binningen, Switzerland
     \and  Observatoire de Besan\c con, 
           41 bis avenue de l'Observatoire, F-25010 Besan\c con, Cedex, France
           }

\offprints{E.Lastennet@qmw.ac.uk\\
$\star$ Based on data from the ESA Hipparcos satellite}

\date{Received 19 February 1999 / Accepted 19 May 1999}

\authorrunning{E. Lastennet {\al}}
\titlerunning{Binaries in the Hyades}

\maketitle

\begin{abstract}
Three binary systems in the Hyades open cluster (51 Tau, 
V818 Tau, and $\theta^2$ Tau), with known \met and 
good Johnson photometric data are used to test the validity of three 
independent sets of stellar evolutionary tracks. 
A statistical method is described and applied to the colour-magnitude 
diagram of the six selected components, giving rise to $\chi^2$-contours 
in the metallicity--age plane. The effects of the Hipparcos parallaxes on 
these confidence regions are studied in detail for these binaries 
through a comparison with  very accurate but older orbital parallaxes. 
There are no significant differences in 51 Tau and $\theta^2$ Tau, and 
hence the test validates the sets of theoretical tracks used. 
However the orbital parallax of V818 Tau leads to metallicities
too large in comparison with the observed range, while the Hipparcos 
parallax give consistent results, at the 1$\sigma$ level. Comparisons 
between the predicted and measured masses are also made and 
show a good agreement. However, even if the Padova tracks predict correct 
masses for each component of V818 Tau, there is no Padova track that fits 
both components simultaneously in the mass-radius diagram.  
Finally, since the individual masses of these binaries are accurately known, 
some issues about the influence of the Hipparcos parallaxes on 
their location in the mass-luminosity diagram are also discussed.  


\end{abstract}

\section{Introduction}

Since the members of an open cluster are assumed to be of same age and chemical 
composition, these stars are currently used to test the validity of  stellar evolution 
theories\footnote{Some of the most recent detailed works on such critical tests 
of stellar evolution in open clusters are due to  Nordstr\"{o}m {\al} (1997) and 
Pols {\al} (1998).}, mainly because main sequence stars define a tight sequence 
in a colour-magnitude diagram (CMD). Unfortunately, this tightness is sometimes
misleading  because of the contamination by field stars, the presence of unresolved binaries 
and also the influence of stellar rotation on the location of massive stars in CMDs. \\
Alternatively, well-detached binaries are powerful tests when fundamental 
parameters are accurately known (see the comprehensive review by Andersen 1991 
on double-lined eclipsing binaries). Unfortunately, the determination of 
the chemical composition often remains a difficult and unresolved issue. 
It appears therefore that  a better test could be performed by combining both 
advantages, that is, testing the tracks with well-detached double-lined binaries 
which are members of open clusters. \\
The work developed in this paper is the application of this idea to three 
well-detached binaries members of the Hyades: 51 Tau, V818 Tau, 
and $\theta^2$ Tau. 
Torres {\al}, 1997 ([TSL97a], [TSL97b] and [TSL97c]) obtained the 
first complete visual-spectroscopic solutions for these systems, from which they 
carefully derived very accurate parallaxes and individual masses. They also 
gathered some individual photometric data in the Johnson system.  
Furthermore, we found useful trigonometric parallaxes information in the 
Hipparcos catalogue (ESA, 1997). By combining the two sources of data, 
we investigate the influence of the Hipparcos parallaxes on our method which was 
developed to test stellar evolutionary models in HR diagrams.  \\
According to Leitherer {\al} (1996), the most widely used stellar 
theoretical tracks in the literature are those computed by the Geneva group 
(Schaller {\al} 1992, Schaerer {\al} 1993a, 1993b, and Charbonnel 
{\al} 1993) and the Padova group (Bressan {\al} 1993, Fagotto {\al} 1994a,b
\footnote{High metallicity tracks (Z $=$ 0.10) are also available for the Padova 
group (Fagotto {\al} 1994c), and the Geneva group (Mowlavi {\al} 1998), 
but they are not relevant for the expected Hyades metallicity.}). 
We also used the stellar tracks from Claret \& Gim\'enez (1992) (CG92  
thereafter). The tests presented in the next section are 
done with these 3 series of stellar tracks to assess the possible
systematics introduced by the  use of a single set.\\
The test we want to perform has two aims. Firstly it has to check whether the two 
components of the systems are on the same isochrone, i.e. on a line defined 
by the same age and the same chemical composition for the two single stars. 
Secondly, as all the selected stars are members of the Hyades  whose \met has been 
well measured, we can also check that the predicted \met from theoretical models 
are correct. Therefore, if one of these two criteria is not clearly fullfilled by 
a given set of tracks, then these models have obvious problems since they do not
account for several observational constraints (namely the \met and/or the photometric 
data).\\ 
We would like to emphasize that we do not claim that the 
6 selected Hyades stars allow us to test without ambiguity any set of theoretical 
stellar tracks. Since the data are presented in colour-magnitude diagrams, 
we are in fact testing not only the validity of the tracks (computed in terms of 
effective temperature and bolometric luminosity in the HR diagram) but also 
the validity of 
the photometric calibrations, and disentangling the relative influence of both is  
a tricky task. 
However, since we use the Basel Stellar Library (BaSeL) photometric calibrations, 
extensively tested and regularly updated for a larger set of parameters (see Lejeune 
{\al} 1997, Lejeune {\al} 1998, Lastennet {\al} 1999 and Lejeune {\al} 1999), we 
assume that the calibrations are reliable for our purposes. 
Moreover, it is worth noticing that in the range of interest  of the  (B$-$V) colour 
in the present work ([0.15, 1.25] mag.), the agreement between the various extant 
calibrations, i.e., the transformations between theoretical $T_{\rm eff}$ and 
observational (B$-$V) colour, is quite 
good around  metallicities close to solar (see, for example, Alonso {\al} 1996 and Fig. 8 
of Flower 1996 for some comparisons). Significant differences do occur at cooler 
temperatures, but they are not relevant for the present study.  
For these reasons, we will assume that the calibrations from the BaSeL 
models are reliable enough for this work.\\
Another crucial point is the question of reddening. The Hyades open cluster does not 
show any evidence for reddening (E(B$-$V)$=$0.003$\pm$0.002, Taylor 1980), consequently 
E(B$-$V) is assumed to be consistent with  0 in this paper. \\ 
The paper is organised as follows: Sect. 2 is a brief description of the method 
used to produce confidence level contours, Sect. 3 is devoted to the study of each 
binary, completed by comments on the influence of the Hipparcos parallaxes on 
the location of the 6 stars in a mass-luminosity diagram. Finally, Sect. 4 draws our 
general conclusions.    

\input{table1.tex}

\section{Statistical method description} 
 
It is not an easy task to disentangle  \met and  age effects in a colour-magnitude 
diagram  with only two stars, particularly if they both lie close to the
main sequence. This difficulty is due to the degeneracy between these two 
quantities in this area of the CMD.  
For this reason, any analysis of binary star data should be presented in the 
(metallicity, age) plane.   
In order to derive simultaneously the \met (Z) and the age (t) of the system, 
we minimize the $\chi^2$-functional defined as: 

\beqa
\chi^2 (t, Z)  & = &  \sum_{i=A}^{B} \left[ \left(\frac{\rm M_V(i)_{\rm mod} -  M_V(i)}
{\sigma(\rm M_V(i))}\right)^2 \right.  \nonumber \\
 &  &+ \left.  \left(\frac{\rm (B-V)(i)_{\rm mod} - (B-V)(i)}{\sigma(\rm (B-V)(i))}\right)^2 \right]
\eeqa
where $A$ is the primary and $B$ the secondary component. M$_V$ and (B$-$V) are the observed 
values (more details will be given later on M$_V$), and M$_V$$_{\rm mod}$ and 
(B$-$V)$_{\rm mod}$ are obtained from the synthetic computations of the BaSeL models
using a given set of stellar tracks. 

A similar method has already been developed and used by Lastennet {\al} (1996) 
and by Lastennet \& Valls-Gabaud (1999) for  
HR diagrams. Equation 1 simply means that the best $\chi^2$ is obtained when an 
isochrone is close to both stars A and B in the CMD. 
With $n=4$ observational data (M$_{\rm V}$ and B$-$V for each star) and  $p=2$ free 
parameters (age t and metallicity Z), we expect to find a $\chi^2$-distribution with 
$q=n-p=2$ degrees of freedom. Finding  the central minimum value $\chi^{2}_{\rm min}$, 
we form  the $\chi^2$-grid in the (metallicity, age)-plane and compute the boundaries 
corresponding to 1$\sigma$, 2$\sigma$, and 3$\sigma$ confidence levels. 
We do not use any additional information in the $\chi^2$ functional such as radii 
or masses since (1) these are not always available at the same time, and (2) they 
should be predicted by the tracks (Lastennet \& Valls-Gabaud 1999).

\section{Tests with well-detached binaries}

In this section tests of the widely used theoretical stellar tracks are presented. They 
are based on a few number of binary systems belonging to the Hyades open cluster, for 
which Johnson BV photometric magnitudes and accurate mass estimates are available: 51 
Tau, V818 Tau, and $\theta^2$ Tau. V818 Tau is the only eclipsing binary among these 
three systems. 
Contrary to Lastennet \& Valls-Gabaud (1999) where tests are presented in the 
theoretical HR diagram, our tests are performed in the CMD. \\
The main advantage to study such binaries, members of an open cluster,  
is that the heavy element abundance of the Hyades has been extensively studied and 
estimated by different authors. As a matter of fact, the \met of the binaries studied 
in Lastennet \& Valls-Gabaud (1999), mainly from the field, was seldom known. 
The more recent determinations of the Hyades \met have been reviewed by Perryman 
{\al} (1998): [Fe/H] $=$ 0.14 $\pm$ 0.05, i.e. Z $=$ $0.024^{+0.0025}_{-0.003}$ assuming 
a subsolar helium abundance (Y $=$ 0.26), which is the value found by Lebreton {\al} (1997) 
in order to reproduce the Hyades main sequence, with an uncertainty of about 0.02 resulting 
both from the error bars of the Hyades members in the HR diagrams and from the observational 
uncertainty on [Fe/H]. 
A second advantage is that very accurate parallaxes are now available from Hipparcos 
for these three systems (the last estimation of the Hyades distance from Perryman {\al} 
(1998) is: d $=$ 46.34 $\pm$ 0.27 pc). Moreover, we pay attention to the quality of the 
Hipparcos results using the indicators F1 (percentage of rejected observations
\footnote{This field gives the percentage of data that had to be rejected in order 
to obtain an acceptable astrometric solution.}) and F2 (goodness-of-fit statistic) 
in order to ensure that the parallaxes are reliable. The data used in the present work 
are gathered in Table 1.

Absolute magnitudes of each binary component are directly inferred from 
Hipparcos parallaxes $\pi$ and visual magnitudes V: 
\begin{equation}
M_V = V + 5 + 5 \times \log (\pi) 
\end{equation}
\begin{equation}
\Delta M_V = \Delta V + \frac{5}{\ln(10)}
\times  \frac{\Delta(\pi)}{\pi}
\end{equation}

We do not take into account the Lutz--Kelker correction because relative 
errors on parallaxes, $\sigma_{\pi}$/$\pi$, are always smaller than 10 \%: 
respectively 4.5 \%, 5.8 \% and 3.8 \% for 51 Tau, V818 Tau and $\theta^2$ Tau.
Therefore the correction can be neglected as noted by Brown {\al} (1997). 
In Table 1, the systematically larger distance found by Torres {\al} is, at 
least in part, caused by the fact that the PPM proper motions that they used 
(see Schwan 1991 and references therein) are almost all smaller than those of 
Hipparcos (see Perryman {\al} 1998 for details).

\subsection{Results for 51 Tau} 

The 51 Tau system is a spectroscopic binary and also a visual binary resolved by 
speckle interferometry.  
\begin{figure*}[htb]
\centerline{\psfig{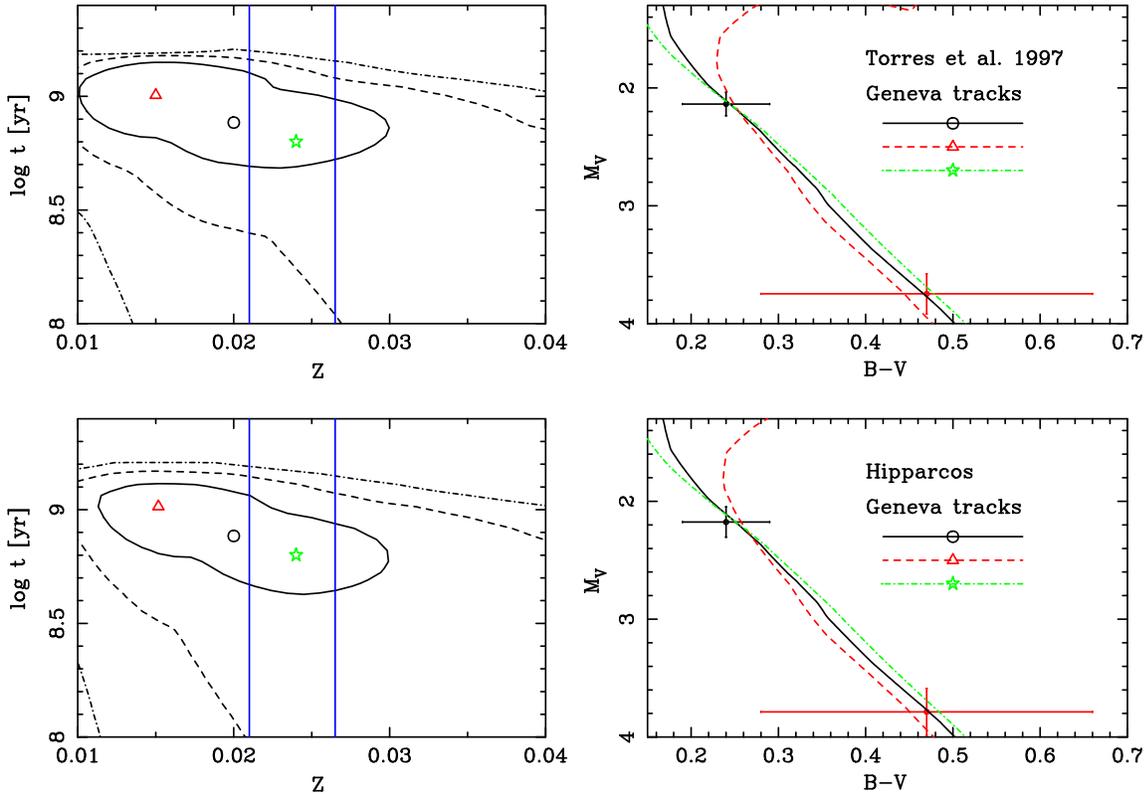}}
\caption{51 Tau system: influence of the Hipparcos parallax on the contour 
levels derived from the Geneva tracks. The photometric locations of each single 
star in the CMD are from Torres {\al} [TSL97a], except the absolute magnitudes 
$M_V$ in the {\it bottom panel} which are derived from the Hipparcos parallax. 
The correspondence between isochrones and symbols are directly given on the 
plots. For instance, the best fit isochrone ({\it solid line} in the CMD) is defined by 
Z $=$ 0.020, log t $=$ 8.88. In isocontours plots, the parameter values corresponding to 
the best fits ($\chi^2_{\rm minimum}$) are marked by {\it open circles}. The metallicity--age pair 
(Z $=$ 0.024, log t $=$ 8.80) from Perryman {\al} (1998) for the Hyades is shown for 
comparison ({\it star} symbol). It is important to notice that all the isochrones inside the 1$\sigma$ 
confidence levels are also good fits: an older isochrone is also a possible fit 
(e.g. the isochrone corresponding to the {\it triangle} defined by Z $=$ 0.015 and log t $=$ 9). 
A comparison between {\it upper and bottom panels} shows that the 1, 2, and 3$\sigma$ contour 
levels (respectively {\it solid}, {\it dashed} and {\it dot-dashed lines}) are not significantly 
modified. Vertical lines in contour levels diagrams show the observational limits for the 
\met of the Hyades.}
\label{f:51Tau}
\end{figure*}
\begin{figure*}[htb]
\centerline{\psfig{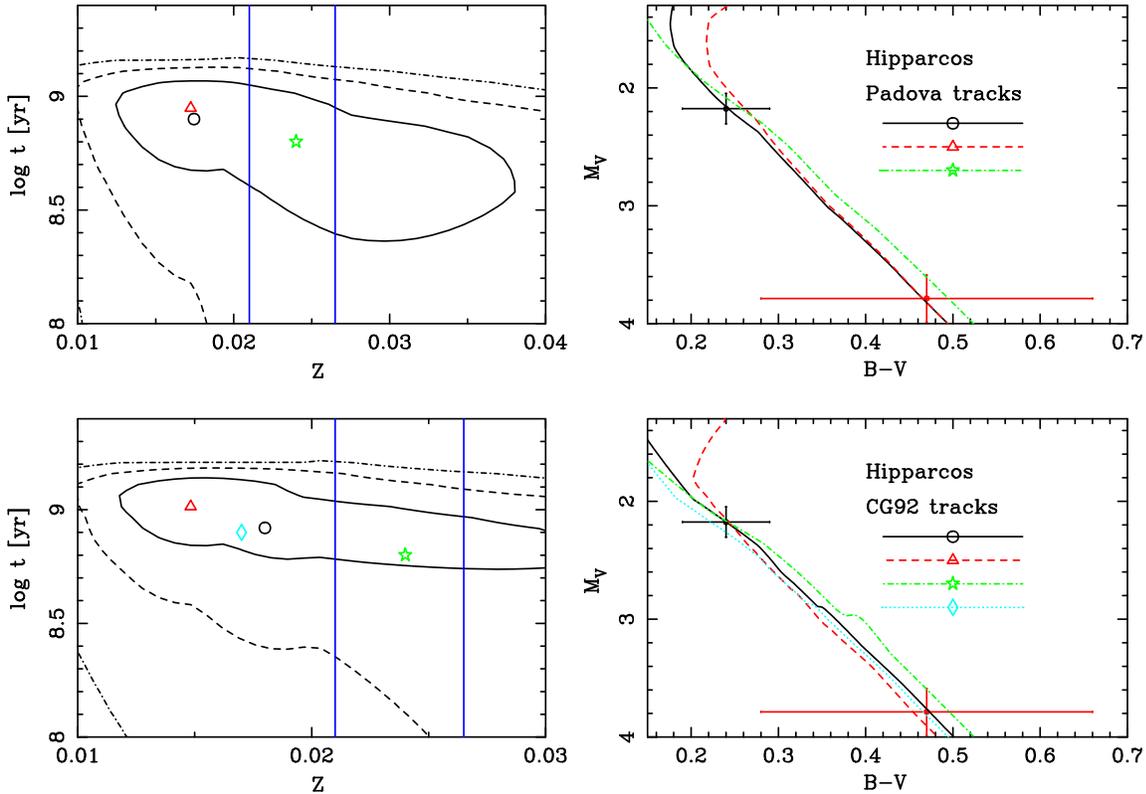}}
\caption{51 Tau system: Padova ({\it upper panels}) and CG92 tracks ({\it lower panels}) 
both taking into account the Hipparcos parallaxes. 
The photometric locations of each star in the CMD are from Torres {\al} [TSL97a], 
but the absolute magnitude $M_V$ is derived from the Hipparcos parallax.
As in Fig. 1, vertical lines in contour levels diagrams show the observational limits of 
the \met of the Hyades, and the correspondence between isochrones and symbols are directly 
given on the plots. The best fits ($\chi^2_{\rm minimum}$) are marked by {\it open 
circles} (Padova: Z $=$ 0.017, log t $=$ 8.90, CG92: Z $=$ 0.018, log t $=$ 8.92), but good fits 
can also be obtained with other isochrones: for example, the isochrones corresponding to the 
{\it star} (log t $=$ 8.80 and Z $=$ 0.024, from Perryman {\al}, 1998) and to the {\it triangle} 
(Padova: Z $=$ 0.017, log t $=$ 8.95, CG92: Z $=$ 0.015, log t $=$ 9.01), slightly older and more 
metal poor than the best solution. In the CG92 diagrams ({\it lower panels}), 
the {\it dotted line} isochrone (Z $=$ 0.017, log t $=$ 8.90) corresponds to the {\it diamond symbol} 
which is very close to the best solution ({\it open circle}). However, the two isochrones are quite 
different. In the CG92 CMD ({\it right bottom panel}), the small kink around B$-$V$=$0.4 
is not due to the colour transformation but to the CG92 model at 1.41$M_{\odot}$.}  
\label{f:51Tau_PC}
\end{figure*}

In order to address the question of the effect of the Hipparcos parallaxes on 
our results, we first need 
to present what  the results are without using these new parallaxes. 
Figure \ref{f:51Tau} shows both the isocontours and CMD without using 
Hipparcos ({\it upper panels}) and with Hipparcos parallaxes ({\it lower 
panels}). The effect of using the Hipparcos parallaxes is hardly visible 
on the results, in particular the shapes of the isocontours are not modified
a great deal.  
Actually, the difference between the distance from Torres {\al} 1997 
(d $=$ 55.8 $\pm$ 1.8 pc from the orbital parallax) and the distance derived from 
Hipparcos (d $=$ $54.8^{+2.6}_{-2.4}$ pc) is rather small. 
Since the errors on  distance are slightly larger with Hipparcos data, the 
error bars estimates are slightly larger on the $M_V$ axis. \\
As mentioned previously, the more recent determinations of the Hyades \met have 
been reviewed by Perryman {\al} (1998): [Fe/H] $=$ 0.14 $\pm$ 0.05 (i.e: 
Z $=$$0.024^{+0.0025}_{-0.003}$). These constraints are shown in the (Z, age)-plane 
as vertical lines. Torres {\al} (TSL97a) adopt [Fe/H]$_{\rm Hyades}$ $=$ 0.13 (following   
 Cayrel {\al} 1985, Boesgaard 1989 and Boesgaard \& Friel 1990) and hence   
 Z $=$ 0.027, which is about the upper limit quoted by  Perryman {\al} (1998). 
From the $\chi^2$-contours in Fig. \ref{f:51Tau} and Fig. \ref{f:51Tau_PC}, one can 
see that the stellar evolutionary tracks from Geneva, Padova and CG92 yield acceptable 
and essentially equivalent fits of isochrones in agreement with the Hyades metallicity. 
However, the preferred solution (i.e. with the minimum $\chi^2$-value) in the sets of 
tracks points systematically to a smaller \met than expected 
(cf. in each Z--age diagram the location of the {\it open circle}, best fit, 
in comparison with the \met  constraints shown with vertical lines). \\ 
Since the masses of the two components of 51 Tau are known to a good accuracy 
(respectively 7\% and 12\%), to check that the predicted masses are in 
agreement with the measured masses is one more stringent test.  
The best Geneva fit sligthly underestimates the masses of 51 Tau (by 0.5$\sigma$ 
for the primary and 1$\sigma$ for the secondary component) as well as the Padova best 
fit (respectively 1$\sigma$ for the primary and $\sim$1.2$\sigma$ for the secondary) 
and CG92 (respectively 0.5$\sigma$ for the primary and 1$\sigma$ for the secondary) 
which means that these models are able to reproduce to a satisfying level the stellar 
masses in the mass range of 51 Tau (see Table 2 for details). 
The agreement is even better if we take into account 
the following point. When spectroscopic and astrometric orbital elements are known, 
one can derive the mass of each component, according to an hypothesis on the distance 
(i.e. the parallax). Hence, if we keep all the orbital parameters (period, eccentricity, 
velocity amplitude, ...) as given in [TSL97a] and take the Hipparcos parallax value of 
51 Tau, one derives (from Eq. 2 in [TSL97c]) the new masses M$_A$$\simeq$1.66$M_{\odot}$ 
and M$_B$$\simeq$1.40$M_{\odot}$, values that are even better matched by the three sets 
of theoretical models. 
\begin{table}[htb] 
\caption[]{Comparison of theoretical mass estimates from isochrone age fitting 
of the system 51 Tau with the values of Torres {\al} [TSL97a].}
\begin{flushleft}
\begin{center}  
\begin{tabular}{ccccc}
\hline\noalign{\smallskip}
 & [TSL97a]  & Geneva$^{a}$ & Padova$^{a}$ & CG92$^{a}$ \\ 
\noalign{\smallskip}
\hline \noalign{\smallskip}
\noalign{\smallskip} 
A & 1.80$\pm$0.13 & 1.74$\pm$0.06 & 1.67$\pm$0.04 & 1.72$\pm$0.05 \\
B & 1.46$\pm$0.18 & 1.27$\pm$0.05 & 1.24$\pm$0.04 & 1.28$\pm$0.05 \\
\noalign{\smallskip}\hline
\end{tabular}
\end{center}
\smallskip
$^{a}$ 1$\sigma$ interval (this work). \\
\end{flushleft}
\end{table}
Assuming the most conservative \met values of the Hyades,  a close inspection of 
the isocontours derived from the Geneva models reveals that the contours provide an age estimate  
-- at the 1$\sigma$ level -- between log t $=$ 8.6 and 9.0 years (i.e. between 0.4 $\times$ 
$10^9$ and $10^9$ years).  
A similar work with the Padova tracks give an age between log t $=$ 8.5 and 9.05,
and with CG92 between log t $=$ 8.75 and 9.05 years. As summarized in Table 3, these 
values are in good agreement with the age estimation of Perryman {\al} (1998): log t 
$=$ $8.80^{+0.02}_{-0.04}$, inferred from the isochrone fitting technique with the CESAM 
stellar evolutionary code (Morel 1997). 
\begin{table}[htb] 
\caption[]{Comparison of theoretical age estimates from isochrone 
age fitting of 51 Tau with Perryman et al's determination.}
\begin{flushleft}
\begin{center}  
\begin{tabular}{ll}
\hline\noalign{\smallskip}
Stellar models & log(age) [yrs] \\ 
\noalign{\smallskip}
\hline\noalign{\smallskip}
\noalign{\smallskip} 
Geneva$^{a}$ &  [8.60, 9.00] \\
Padova$^{a}$ &  [8.50, 9.05] \\
CG92$^{a}$   &  [8.75, 9.05] \\
CESAM$^{b}$  &  [8.76, 8.82] \\
\noalign{\smallskip}\hline
\end{tabular}
\end{center}
\smallskip
$^{a}$ 1$\sigma$ interval (this work), assuming  
0.021$<$$Z_{\rm Hyades}$$<$0.026. \\
$^{b}$  Perryman {\al} (1998). \\
\end{flushleft}
\end{table}

\begin{figure*}[htb]
\centerline{\psfig{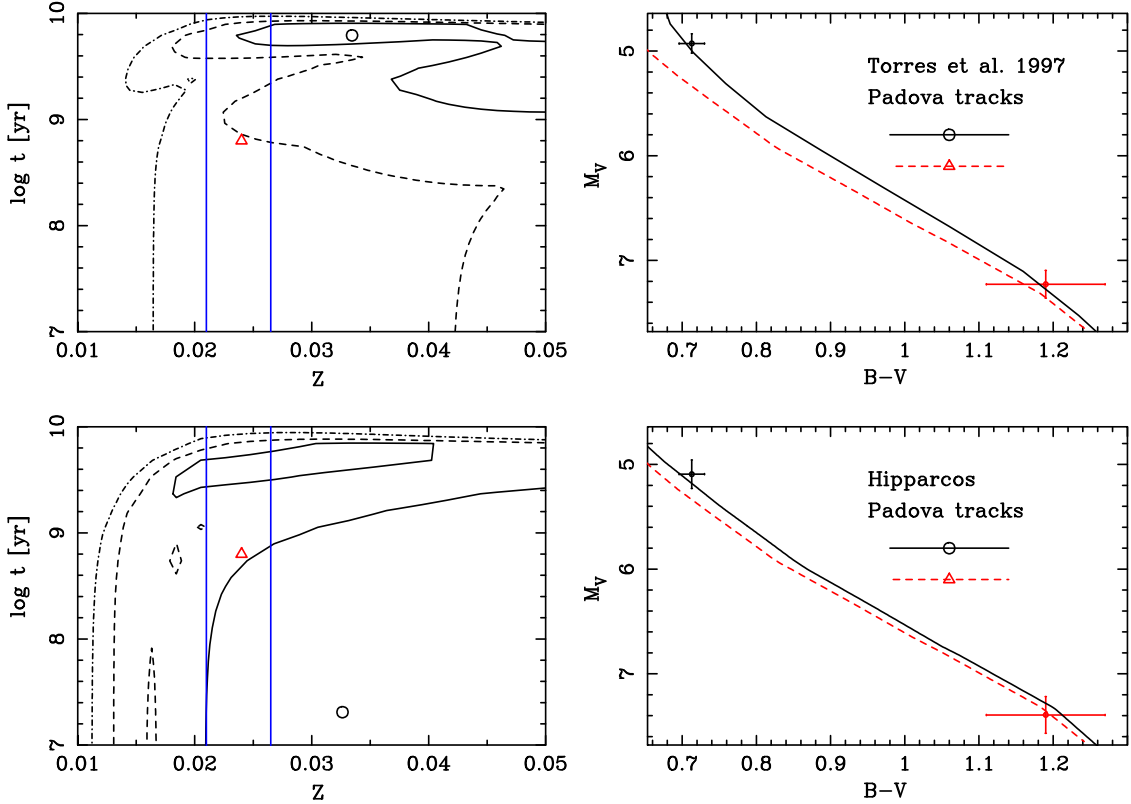}}
\caption{V818 Tau system: influence of the Hipparcos parallaxes on the contour 
levels derived from the Padova tracks. The photometric locations of each star in the 
CMD are from Torres {\al} [TSL97a], except the absolute magnitude $M_V$ in the 
{\it bottom panel} which is derived from the Hipparcos parallax. 
The best fit isochrones ({\it solid lines}) are defined by Z $=$ 0.033, log t $=$ 9.79
(with Torres {\al} parallaxes, {\it upper CMD}) and Z $=$ 0.033, log t $=$ 7.30 (Hipparcos, 
{\it bottom CMD}). It is crucial to notice that all the isochrones inside the 1$\sigma$ confidence 
levels are also good fits. 
Vertical lines in contour levels diagrams show the observational limits for the \met of the Hyades. 
The metallicity--age couple (Z $=$ 0.024, log t $=$ 8.80) from Perryman {\al} (1998) for 
the Hyades is also shown for comparison ({\it triangle}). The corresponding isochrone 
({\it dashed line}) clearly does not fit the system.  
}
\label{f:V818Tau}
\end{figure*}
Finally, we would like to point out the need of new photometric data: the uncertainty 
on the Johnson (B$-$V) colour of the faintest star is larger by a factor of 4 than 
the primary. So, even if it appears clearly that the three sets of tracks do not fail 
the test, accurate (B$-$V) data for the less massive component at the same level as 
the primary should reduce the extent of the contours, allowing to perform more 
stringent tests.

\subsection{Results for V818 Tau}

\begin{figure*}[htb]
\centerline{\psfig{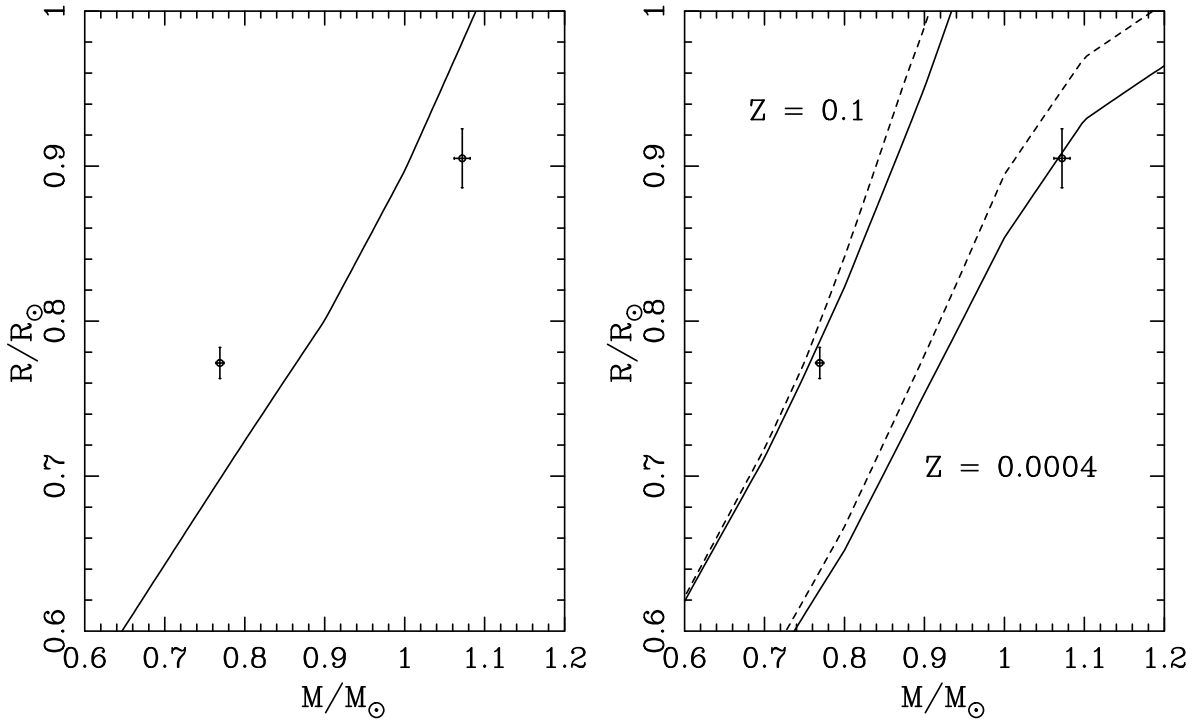}}
\caption{V818 Tau system in a Mass-Radius diagram where all the isochrones are derived from 
the Padova tracks. 
The mass of each star is from Peterson \& Solensky (1988), as quoted by Torres {\al} [TSL97a], 
and the radius is from Schiller \& Milone (1987). 
In the {\it left panel}, the {\it solid line} isochrone (log t $\simeq$ 7.30, Z $\simeq$ 0.033) 
is the best fit isochrone shown in the CMD of Fig. 3. Even if this isochrone allows to predict the 
correct mass for each component, it clearly does not fit V818 Tau in the mass-radius diagram. 
In the {\it right panel}, four different isochrones are shown for comparison: with a high 
metallicity (Z $=$ 0.1) for two ages, t $=$ 0 ({\it solid line}) and t $=$ $10^9$ yrs ({\it dashed 
line}) and with a very low metallicity (Z $=$ 0.0004) for the ages t $=$ 0 ({\it solid line}) and 
t $=$ $10^9$ yrs ({\it dashed line}). There is no Padova isochrone which is able to fit 
simultaneously both components of V818 Tau in the mass-radius diagram.  
}
\label{f:V818_mr}
\end{figure*}
\begin{figure*}[htb]
\centerline{\psfig{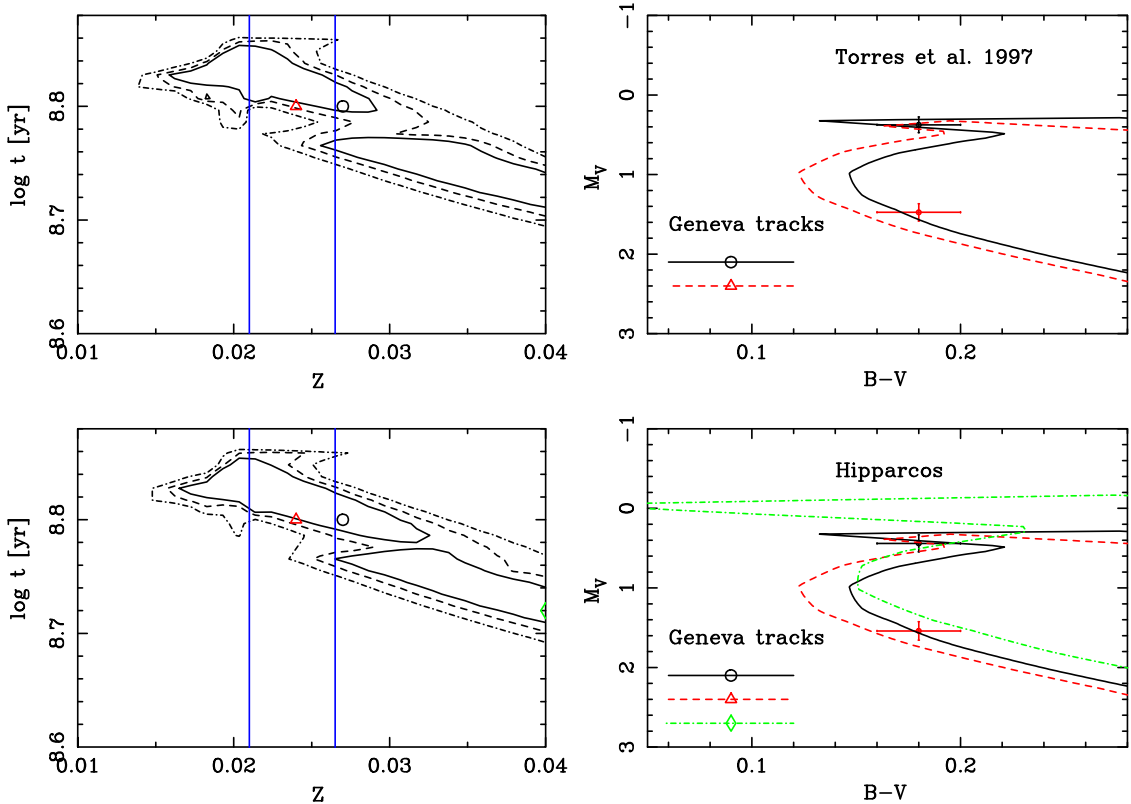}}
\caption{$\theta^2$ Tau system: influence of the Hipparcos parallaxes on the contour 
levels derived from the Geneva tracks. The photometric locations of each star in the 
CMD are from Torres {\al} [TSL97a], except the absolute magnitude $M_V$ in the {\it 
bottom panel} which is derived from the Hipparcos parallax. Vertical lines in contour 
levels diagrams show the observational limits of the \met of the Hyades. {\it Solid 
lines} isochrones in the CMD correspond to the best $\chi^2_{min}$ fits ({\it open 
circles}, Z $=$ 0.027 and log t $=$ 8.80) and {\it dashed lines} isochrones correspond to the 
{\it triangle} (Z $=$ 0.024 and log t $=$ 8.80, values from Perryman {\al} 1998 for the Hyades). 
Another isochrone ({\it dot-dashed line}) corresponding to the {\it diamond} symbol 
(Z $=$ 0.04 and log t $=$ 8.72) in the (t,Z) diagram is also shown in the Hipparcos CMD. 
This solution is still an acceptable CMD fit at the 1$\sigma$ confidence level 
even if the \met is twice solar,  which is excluded by observational constraints.}
\label{f:Theta02Tau}
\end{figure*}
The V818 Tau system (vB22) is a double-lined eclipsing binary (Mc Clure, 1982) with very 
well estimated masses (actually the most accurate masses known for Hyades members). 
Indeed, the relative errors on the masses are less than 1\%, and the secondary component 
is particularly interesting because it is one of the rare stars which is less massive than 
the Sun and whose mass is known with such high accuracy (cf. Andersen {\al} 1991). Unfortunately, 
this low mass star (about 0.77 $M_{\odot}$) does not allow us to test either the CG92 models 
-- the lower available mass of these models being 1 $M_{\odot}$ -- or the Geneva 
models whose lower mass limit is 0.8 $M_{\odot}$. Nevertheless, we have tested Geneva 
tracks (for the primary component) coupled\footnote{This simply means that the Geneva 
tracks were extrapolated with the Baraffe {\al} models in the low mass region 
(below 0.8 $M_{\odot}$) for each metallicity. Of course, this point is a very rough 
approximation because the physics of these models are different, but it will not 
change our final conclusions.}  with the  Baraffe {\al} (1995) 
models (for the low-mass secondary star), as well as the Padova group models whose 
lower mass limit is small enough for our purpose (0.6 $M_{\odot}$).\\
Torres {\al} [TSL97a] derived for this binary the distance d $=$ 50.4 $\pm$ 1.9 pc  
indicating that V818 Tau is slightly more distant than the cluster center.  
By using the Hipparcos parallax, we obtain d $=$ $46.7^{+2.9}_{-2.6}$ pc, 
which seems to be in a better agreement with the distance of the Hyades center.  
The confidence levels that we obtain by combining pre-Hipparcos data (see {\it upper panels} 
in Fig. \ref{f:V818Tau}) and the isochrone fitting technique with the Padova tracks in 
order to match simultaneously the locations of the two stars in the CMD, are in a 
clear disagreement with the observational constraints of the Hyades metallicity. 
As a matter of fact, the Padova as well as the Geneva tracks 
predict metallicities much larger than expected from observational determinations: 
1.5 to 2 times solar metallicities (Z$\in$[0.03,0.04]) provide the best fits as shown 
in the {\it upper panels} of Fig. \ref{f:V818Tau}. The Padova isochrone corresponding to the 
metallicity--age Hyades solution of the detailed work of Perryman {\al} (1998) is 
also shown ({\it dashed line} in the upper CMD of  Fig. \ref{f:V818Tau}). This solution should 
be consistent with the observational \met of the Hyades, however, the fit is clearly 
bad in the CMD.\\  
Now, if we consider the Hipparcos parallax of this binary system, the absolute 
magnitude (see in particular the primary component, which puts the more stringent 
constraints) is shifted towards fainter magnitudes, giving rise to significantly different 
$\chi^2$-contours (see confidence contours in the {\it lower panel} of Fig. \ref{f:V818Tau}). 
Hipparcos parallaxes have a strong influence on the shape of the isocontours 
that we obtain: for instance, 1$\sigma$ confidence levels occupy a larger region with 
Hipparcos and are in a better agreement with the observational constraints 
({\it vertical lines} in the Z-t diagram). 
We wish to emphazise that even if the best displayed Z-value inferred from Hipparcos 
({\it open circle} in the {\it bottom panel} of Fig. \ref{f:V818Tau}) is actually the same as with 
Torres {\al} (1997) (Z $\simeq$ 0.033), it is worth noticing that the Hipparcos parallaxes give also 
possible solutions at 1$\sigma$ level with metallicities and ages lower than in the 
pre-Hipparcos Z--t diagram ({\it left upper panel}) where solutions are confined in a  
region roughly defined by Z $\geq$ 0.024 if log t $\geq$ 9.7, and Z $\geq$ 0.037 if 
log t $\geq$ 9. 
Therefore, if one considers the confidence levels, the disagreement found using orbital 
parallaxes disappears, and the models do not fail anymore. 
We would like to mention the fact that this result is rather puzzling because 
the orbital parallaxes are a little more accurate than the Hipparcos parallaxes, that 
is why such variation was not expected. The change of shape is due to the fact that 
the distance derived from Hipparcos for V818 Tau is about 8\% smaller than from its 
orbital parallax. The better consistency between the contours diagram of the Hipparcos 
solution and the \met of the Hyades, which is the only reliable 
constraint, gives more weight to the Hipparcos distance than to the orbital parallax.  
This example highlights the importance of having very accurate but also reliable 
fundamental data, because small errors on the data can lead to reject -- by 
mistake -- the validity of a stellar model. \\
As mentioned in the beginning of this section, the masses of the two components of V818 Tau 
are known with an excellent accuracy (better than 1\%, following the study of Peterson 
\& Solensky, 1988, [PS88]). Therefore, to check that the 
stellar masses predicted by the models are in agreement with the true masses is 
probably the most critical test. A comparison is given in Table 4. 
The result is that the best Padova fit underestimates 
the masses of V818 Tau by about 5$\sigma$ for both stars. This is not so bad because 
of the high level of accuracy of the masses: 5$\sigma$ only means 0.05$M_{\odot}$ 
for V818 Tau A and 0.025$M_{\odot}$ for V818 Tau B. \\ 
\begin{table}[htb] 
\caption[]{Comparison of theoretical mass estimates from Padova isochrone age fitting 
of the system V818 Tau with the values of [SM87] (Schiller \& Milone, 1987) and [PS88] 
(Peterson \& Solensky, 1988).}
\begin{flushleft}
\begin{center}  
\begin{tabular}{cccc}
\hline\noalign{\smallskip}
  & [SM87] & [PS88]  & Padova$^{a}$ \\ 
\noalign{\smallskip}
\hline\noalign{\smallskip}
\noalign{\smallskip} 
A & 1.080$\pm$0.017 & 1.072$\pm$0.010 & 1.020$\pm$0.04  \\
B & 0.771$\pm$0.011 & 0.769$\pm$0.005 & 0.744$\pm$0.02  \\
\noalign{\smallskip}\hline
\end{tabular}
\end{center}
\smallskip
$^{a}$ 1$\sigma$ interval (this work). \\
\end{flushleft}
\end{table}
Moreover, since V818 Tau is a double-lined eclipsing binary, the radius of each component 
is also known with an excellent accuracy (better than 2\%). The stellar radius can also be 
computed from the effective temperature and the luminosity given by the Padova models. 
Therefore, it is possible to compare the stellar radii predicted by the models with the true 
radii. Unfortunately, even if Padova models predict a correct mass for each component 
of V818 Tau, it appears clearly from Fig. \ref{f:V818_mr} that there is no Padova isochrone able 
to fit both components of V818 Tau in the mass-radius diagram. Even with the best fit isochrone 
obtained from the CMD of Fig. 3, the radius of the more massive component is overestimated 
by more than 0.3 $R_{\odot}$, and the radius of the less massive component is underestimated 
by more than 0.3 $R_{\odot}$. 
This test shows that masses are much less discriminant, as far as the power of the test is 
concerned, than radii, hence the importance of double-lined eclipsing binaries to fully 
constrain stellar tracks.

\subsection{Results for $\theta^2$ Tau}

The primary component of the spectroscopic binary $\theta^2$ Tau (spectral type A7 III) is 
one of the brightest stars in the Hyades. 
Such a system is {\it a priori} very critical because it is composed by a main-sequence star 
and an  evolved star, which allows us to test widely different evolutionary stages.
The tests that we performed with this binary are shown in Fig. \ref{f:Theta02Tau} 
where the locations in the CMD are from Torres {\al} [TSL97c], except in the 
{\it bottom panel} where the absolute magnitude $M_V$ is derived from the Hipparcos 
parallax. \\
The first point is that the Hipparcos mission parallax has hardly an influence 
on the luminosity of the two components. Therefore, the 1, 2 and 3$\sigma$ $\chi^2$-contour  
levels show no real significant differences. Moreover, the best fits obtained in 
Fig. \ref{f:Theta02Tau} present a very similar age and metallicity, putting the most massive 
component at the end of the core hydrogen burning phase as pointed out by Torres {\al}  
[TSL97c]. Nevertheless, if one believes a \met slightly poorer (for instance the solution 
shown as a triangle symbol in the {\it lower panel} of Fig. \ref{f:Theta02Tau}) -- more in 
agreement with the Hyades \met and still acceptable inside the 1$\sigma$-contour levels -- 
then this massive star could already be on the begining of the hydrogen-shell burning phase. \\
It is also worth noticing that both components of $\theta^2$ Tau (also known as 78 Tau) 
are fast rotators. The primary component has v $\sin$ i $\simeq$ 80 \kms 
(Slettebak {\al} 1975) and the secondary seems to rotate even faster (estimations range 
from 90 to 170 \kms). Therefore, the rotational effect on their  locations 
in the CMD  should certainly not be neglected. The effect expected is typically a few
tenths of magnitude in absolute magnitude and a few hundredths of magnitude in B$-$V, but 
these values are highly dependent of spectral type, age, and chemical composition (see for 
instance Maeder 1971 and Zorec 1992). As a direct consequence, this effect could modify the 
results presented in Fig. \ref{f:Theta02Tau} because they are inferred from theoretical 
tracks which neglect the stellar rotation effect. However, a detailed study of this point 
is beyond the scope of this paper (and will be developed in a forthcoming paper). \\
Nevertheless, in the particular case of $\theta^2$ Tau, the stellar rotation effect 
should not call into question the validity of the theoretical tracks of Geneva, Padova 
or CG92, because the contours show a large range of possible \met (isocontours derived 
from Padova models have the same shape/are identical than those of Geneva, as well 
as the CG92 models whose upper limit is Z$=$0.03). Then, even with slight variations of 
magnitude and colour in the CMD, a part of these $\chi^2$-contours will always be 
in agreement with the observed \met contraints. \\
\begin{table}[htb] 
\caption[]{Comparison of theoretical mass estimates from Geneva isochrone age fitting 
of the system $\theta^2$ Tau with the values of [TPM95] (Tomkin {\al}, 1995) 
and [TSL97c] (Torres {\al}, 1997).}
\begin{flushleft}
\begin{center}  
\begin{tabular}{cccc}
\hline\noalign{\smallskip}
  & [TPM95] & [TSL97c]  & Geneva$^{a}$ \\ 
\noalign{\smallskip}
\hline\noalign{\smallskip}
\noalign{\smallskip} 
A & 2.1$\pm$0.3 & 2.42$\pm$0.30 & 2.37$\pm$0.02  \\
B & 1.6$\pm$0.2 & 2.11$\pm$0.17 & 1.95$^{+0.06}_{-0.03}$  \\
\noalign{\smallskip}\hline
\end{tabular}
\end{center}
\smallskip
$^{a}$ 1$\sigma$ interval (this work). \\
\end{flushleft}
\end{table}
Since the masses of the two components of $\theta^2$ Tau are known to a good accuracy 
(respectively 12\% and 8\%), to check that the predicted masses are in 
agreement with the measured masses is one more stringent test. 
Torres {\al} [TSL97c] updated the previous work on this system (Tomkin {\al} 1995), 
which resulted in an increase of the masses ($M_A$ $=$ 2.42 $\pm$ 0.30 $M_{\odot}$ instead of 
$M_A$ $=$ 2.10 $\pm$ 0.60 $M_{\odot}$ and $M_B$ $=$ 2.11 $\pm$ 0.17 $M_{\odot}$ instead of 
$M_B$ $=$ 1.60 $\pm$ 0.40 $M_{\odot}$).
As shown in Table 5, the best Geneva fit gives a quite 
good agreement  which means that these models are able to reproduce to a satisfying 
level the stellar masses in the mass range of $\theta^2$ Tau. \\
The general conclusion of the study performed with $\theta^2$ Tau is that the three 
theoretical models allow us  to fit correctly the system for (metallicity, age)-pairs 
in agreement with the more recent constraints available about the \met of the Hyades 
cluster. 

\subsection{The influence of the Hipparcos parallaxes on the Hyades mass-M$_V$ relation}

\begin{figure*}[htb]
\centerline{\psfig{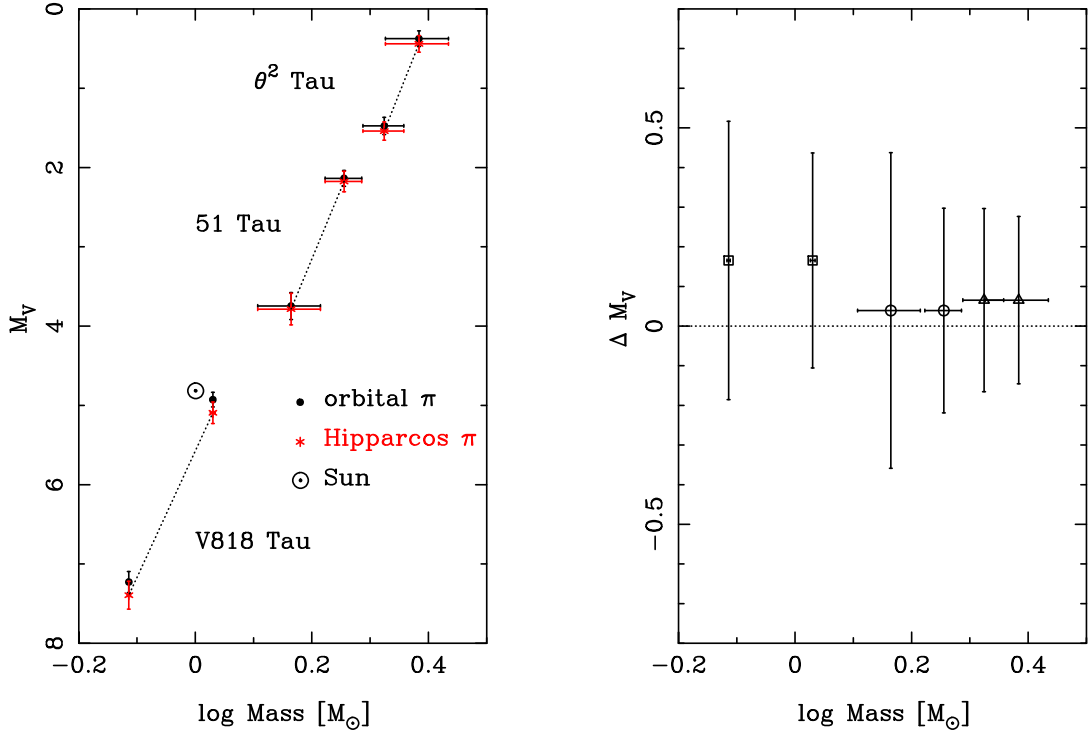}}
\caption{Influence of the Hipparcos parallaxes on the Hyades mass-M$_V$ 
relation of the 6 stars studied in the present work. The components of
each  binary system are 
connected by {\it dotted lines} and the Sun is also shown for comparison 
(M$_V$$=$4.83 from Bessell 1991). On the {\it right panel}, V818 Tau ({\it square}), 
51 Tau ({\it circle}) and $\theta^2$ Tau ({\it triangle}) are displayed on 
a mass-$\Delta$M$_V$ diagram, where $\Delta$M$_V$ means M$_V$ from Hipparcos 
parallaxes minus M$_V$ from orbital parallaxes. The $\Delta$M$_V$$=$0 line 
({\it dotted line}) is shown for comparison.}
\label{f:ml}
\end{figure*}
The discussion of the empirical mass-luminosity relation of the Hyades is beyond 
the scope of this work, because we only studied a tiny sample of member stars 
(see Torres {\al} 1997 [TSL97a,b,c] for a critical and detailed discussion). However, it is 
interesting to show the influence of the Hipparcos parallaxes on the 
location of our working sample in a mass-luminosity diagram. 
As it is only the influence of Hipparcos that is under scrutiny,  one can 
conclude from Fig. \ref{f:ml} ({\it left panel}) that the global shape of this 
empirical relation does not change. Actually, such a result was expected because 
of the excellent accuracy of the orbital parallaxes obtained before Hipparcos, 
but it was important to check this point. 
Thus, the difference, $\Delta$M$_V$ (M$_V$ from Hipparcos parallaxes minus 
M$_V$ from orbital parallaxes) is always smaller than 0.2 mag. and the error bars 
show a very good agreement with $\Delta$M$_V$$=$0 (Fig. \ref{f:ml}, {\it right panel}). \\

\section{Conclusion}

The stellar theoretical models and tracks from Geneva, Padova and CG92 have been tested 
with well-detached binary systems in the Hyades open cluster: 51 Tau, V818 Tau, 
and $\theta^2$ Tau. Firstly, for each of the 3 selected systems, we computed 
simultaneous metallicity-age solutions to fit the two components in the CMD 
(a summary of the results is given in Table 6, taking into account the Hipparcos parallax).
These results are shown in confidence regions in the (Z,t)-plane. 
Under the reasonable assumption that the metallicity must be the same and equal 
to the Hyades \met for the 6 components, we had one more constraint, seldomly available 
for field binaries. Moreover, the new Hipparcos parallaxes that we found for these 
systems allowed us to put more weight on the power of the test. \\
\begin{table*}[htb] \caption[]{Summary of the theoretical simultaneous metallicity--age estimates
obtained from isochrone age fitting (1$\sigma$ level, this work) taking into account the Hipparcos
parallax.}  
\begin{flushleft} 
\begin{center} 
\begin{tabular}{lllllll} 
\hline\noalign{\smallskip} 
System         & \multicolumn{2}{c}{Geneva} & \multicolumn{2}{c}{Padova} & \multicolumn{2}{c}{CG92} \\
\noalign{\smallskip}
\hline\noalign{\smallskip}
               & Z & log t                  & Z & log t                   & Z & log t               \\ 
\noalign{\smallskip} \hline\noalign{\smallskip}
\noalign{\smallskip}
51 Tau         & 0.020$^{+0.010}_{-0.008}$ & 8.88$^{+0.22}_{-0.23}$ & 
0.017$^{+0.021}_{-0.005}$ & 8.90$^{+0.15}_{-0.55}$ & 
0.018$^{+0.012}_{-0.006}$ & 8.92$^{+0.23}_{-0.17}$ \\ 
\noalign{\smallskip}
V818 Tau$^{a}$ &                           &                        & 
0.033$^{+0.017}_{-0.015}$ & 7.30$^{+2.50}_{-0.30}$ & 
                          &                        \\ 
\noalign{\smallskip}
$\theta^2$ Tau & 0.027$^{+0.013}_{-0.010}$ & 8.80$^{+0.05}_{-0.09}$ & 
0.027$^{+0.023}_{-0.011}$ & 8.80$^{+0.03}_{-0.11}$ &
0.027$^{+0.003}_{-0.005}$ & 8.88$^{+0.02}_{-0.02}$ \\ 
\noalign{\smallskip}\hline \noalign{\smallskip} \end{tabular}
\end{center} 
\smallskip
$^{a}$ As explained in Sect.3.2, the Geneva and CG92 models can not be tested with
the less massive component of V818 Tau.  
\end{flushleft}
\end{table*}
The results of these tests are very satisfying for the 3 sets of tracks, except  
for the case of V818 Tau if we used the orbital parallax (unfortunately, the CG92 
and Geneva models can not be tested with the V818 Tau data because the mass of the
secondary is not in the proper range). 
As shown by the 1$\sigma$ confidence levels, the Padova tracks predict a slightly 
too rich \met for the V818 Tau members. The Geneva models also predict a larger \met 
but we wish to emphazise the fact that in this particular case the test 
does not have the same pertinence because we combined the Geneva with the Baraffe 
{\al} (1995) models in order to be able to fit the low mass secondary component of 
V818 Tau. Nevertheless, even if the best $\chi^2$-fit still gives a too 
large metallicity (reported as the central value in Tab. 6: Z$\simeq$0.033) with Hipparcos, 
the inconsistency with the Hyades \met disappears as soon as we use the Hipparcos parallax, 
because the 1$\sigma$ confidence levels are in agreement with the \met constraints.\\
We also present some comparisons between the masses predicted by the models and 
the true stellar masses which give more weight to the quality of the models. \\
Nevertheless, although the Padova models predict a correct mass for each component 
of V818 Tau, there is no Padova isochrone able to fit both components of V818 Tau 
in the mass-radius diagram. \\
As explained above, these tests are not completely unambiguous because
of the inherent problem of calibration between colour and effective temperature. 
However we do not expect significant errors due to the photometric transformations 
because we use the reliable transformations from the BaSeL models in the well known 
Johnson photometric system and also because no disagreement exists
in the range of (B$-$V) colour relevant for the present study, for a \met close to solar. 
Finally, for illustration purposes, we also present the influence of the Hipparcos 
parallaxes on the empirical mass-luminosity relation existing for these 6 stars. 

\begin{acknowledgements}  
  E.L. thanks F. Genova (Director, CDS, Strasbourg Observatory) for useful suggestions 
  on a preliminary version of this work and J. Fernandes for his comments. 
  We acknowledge an anonymous referee for helpful suggestions which have 
  improved the clarity of this paper. 
  E.L. acknowledges support from the CDS  
  and PPARC under the QMW Astronomy research Grant ZIB3. 
  T.L. gratefully acknowledges support through the Swiss National Science 
  Foundation (grant 20-53660.98 to Prof. R. Buser). 
  This research has made use of the CDS resource facilities 
  and the Simbad database operated at CDS, Strasbourg, France. 
\end{acknowledgements}


\end{document}

%% file: table1.tex
\begin{table*}[htb] 
\caption[]{Cross identification and basic data for the 3 selected binary systems in the Hyades. 
Distances are derived quantities from either orbital parallaxes (Torres {\al} 1997 [TSL97a]) 
or Hipparcos trigonometric parallaxes. Masses and Johnson photometry are from Torres {\al} (1997) [TSL97a].}
\begin{flushleft}
\begin{center}  
\begin{tabular}{lrrrccccc}
\hline\noalign{\smallskip}
 Stars       & HD & HIP & vB$^{\dag}$ & Mass          &  V  &  B$-$V & Distance (pc)      &  Distance (pc)    \\
             &    &     &             & ($M_{\odot}$) &     &        & Torres {\al} (1997) &  Hipparcos \\
\noalign{\smallskip}
\hline\noalign{\smallskip}
\noalign{\smallskip}
51 Tau A         & 27176 & 20087 & 24  & 1.80 $\pm$ 0.13         &  5.87 $\pm$ 0.03 & 0.24 $\pm$ 0.05   & 55.8 $\pm$ 1.8       & 54.8 $\pm$ 2.5  \\ 
51 Tau B         & ...   & ...   & ... & 1.46 $\pm$ 0.18         &  7.48 $\pm$ 0.10 & 0.47 $\pm$ 0.19   & ...                  & ...             \\
V818 Tau A       & 27130 & 20019 & 22  & 1.072 $\pm$ 0.010$^{a}$ &  8.44 $\pm$ 0.01 & 0.713 $\pm$ 0.017 & 50.4 $\pm$ 1.9       & 46.7 $\pm$ 2.7  \\
V818 Tau B       & ...   & ...   & ... & 0.769 $\pm$ 0.005$^{a}$ & 10.74 $\pm$ 0.05 & 1.190 $\pm$ 0.08  & ...                  & ...             \\
$\theta^2$ Tau A & 28319 & 20894 & 72  & 2.42 $\pm$ 0.30$^{b1}$  &  3.74 $\pm$ 0.02 & 0.18 $\pm$ 0.02   & 47.1 $\pm$ 1.7$^{c}$ & 45.7 $\pm$ 1.8  \\
$\theta^2$ Tau B & ...   & ...   & ... & 2.11 $\pm$ 0.17$^{b2}$  &  4.84 $\pm$ 0.03 & 0.18 $\pm$ 0.02   & ...                  & ...             \\
\noalign{\smallskip}\hline
\end{tabular}
\end{center}
$^{\dag}$ The acronym vB stands for van Bueren.\\
$^{a}$ from Peterson \& Solensky (1988). \\
$^{b1}$ from [TSL97c]. The value originally quoted by [TSL97a], 2.10 $\pm$ 0.60 $M_{\odot}$, 
is the determination of Tomkin {\al} (1995) adjusting the error upward by a 
factor of two. \\
$^{b2}$ from [TSL97c]. As $^{b1}$, the value quoted by [TSL97a], 1.60 $\pm$ 0.40 $M_{\odot}$, 
is also from Tomkin {\al} (1995) adjusting the error upward by a factor of two.\\
$^{c}$ from [TSL97c]. Other values: 44.1$\pm$1.7 pc in [TSL97a], 44.1$\pm$2.2 in Tomkin {\al} (1995) and 
42.4$\pm$1.5 pc in Peterson {\al} (1993). \\
\end{flushleft}
\end{table*}